\documentclass[journal]{IEEEtran}
\usepackage{amsmath}
\usepackage{amssymb}
\usepackage{graphicx}
\usepackage{setspace}
\usepackage{xcolor}
\usepackage{url}
\usepackage{enumerate}
\usepackage{mathdots}
\usepackage{epstopdf}
\usepackage[dvips]{epsfig}
\usepackage{fancyhdr}
\usepackage{amssymb}

\newcommand{\ts}{&\hspace{-0.07in}}
\newcommand{\sign}{{\rm sign}}
\newcommand{\nn}{\nonumber}
\newcommand{\beq}{\begin{equation}}
\newcommand{\eeq}{\end{equation}}
\newcommand{\bea}{\begin{eqnarray}}
\newcommand{\eea}{\end{eqnarray}}
\newcommand{\bef}{\begin{flalign}}
\newcommand{\eef}{\end{flalign}}

\newtheorem{theorem}{\textit Theorem}
\newtheorem{lemma}{\textit Lemma}
\newtheorem{remark}{\textit Remark}

\begin{document}

\title{Piecewise Constant Tuning Gain Based Singularity-Free MRAC with Application to Aircraft Control Systems}
\author{Zhipeng~Zhang, Yanjun~Zhang, \IEEEmembership{Member, IEEE}, and Jian~Sun, \IEEEmembership{Senior Member, IEEE}
	\thanks{This work was supported in part by the National Natural Science Foundation of China under Grants 62322304,
		61925303, 62173323, 62003277, 62088101, U20B2073, in part by the Foundation under Grant 2019-JCJQ-ZD-049,
		and in part by Beijing Institute of Technology Research Fund Program for Young Scholars. (Corresponding author: Yanjun Zhang.) }
	\thanks{Zhipeng Zhang, Yanjun Zhang and Jian Sun are with the State Key Lab of Autonomous Intelligent Unmanned Systems and School of Automation, Beijing Institute of Technology, Beijing 100081, China, and Jian Sun also with the Beijing Institute of Technology Chongqing Innovation
		Center, Chongqing 401120, China (Emails: zhipengzhang\_bit@163.com; yanjun@bit.edu.cn; sunjian@bit.edu.cn). }
}

	

\maketitle
\thispagestyle{fancy}            
\fancyhead{}                     
\lhead{This work has been submitted to the IEEE for possible publication. Copyright may be transferred without notice, after which this version may no longer be accessible.}                
\cfoot{\quad}     

\renewcommand{\headrulewidth}{0pt}      
\renewcommand{\footrulewidth}{0pt}

\begin{abstract}
This paper introduces an innovative singularity-free output feedback model reference adaptive control (MRAC)
 method  applicable to a wide range of continuous-time linear time-invariant (LTI) systems with general
 relative degrees. Unlike existing solutions such as Nussbaum and multiple-model-based methods, which manage
 unknown high-frequency gains through persistent switching and repeated parameter estimation, the proposed
 method circumvents these issues without prior knowledge of  the high-frequency gain or additional design
 conditions.
The key innovation of this method lies in transforming the estimation error equation into a linear
 regression form via a modified MRAC law with a piecewise constant tuning gain developed in this work. This
  represents a significant departure from existing MRAC systems, where the estimation error equation is
   typically in a bilinear regression form. The linear regression form facilitates the direct estimation of
    all unknown parameters, thereby simplifying the adaptive control process.
The proposed method preserves closed-loop stability and ensures asymptotic output tracking, overcoming some
 of the limitations associated with existing methods like Nussbaum and multiple-model based methods. The
  practical efficacy of the developed MRAC method  is demonstrated through detailed simulation results
   within an aircraft control system scenario.
\end{abstract}



\section{Introduction}
With the increasing demand for robustness and control accuracy of control systems in manufacturing,
aerospace and other fields, adaptive control has attracted extensive attention from researchers because of
its remarkable adaptability to system uncertainties.
So far, adaptive control has made significant progress, with many valuable research results being reported (\cite{sa89,kk95,is96,zc96,ak08,ll11,ek13,gs14,hw23,wk23,mg24}).

In recent decades, the field of control systems has seen rapid advancements in model reference adaptive control~(MRAC) (\cite{is96,bd98,t03,nn18,zz22,yb24}).
In the traditional MRAC framework, a key feature is the coupling between
the unknown high-frequency gain and the unknown estimation error associated with the system parameters,
from which the estimation error equation caused by this interaction takes on a form of bilinear regression.
To address this issue, the majority of output feedback MRAC methods,
such as the renowned augmented error-based method
(\cite{is96,t03}),
choose a strategy that incorporate the high-frequency gain information.
This information is utilized to
design parameter update laws and prevent the singularity
of MRAC laws. However, these design constraints on the
high-frequency gain hinder the further development of
MRAC and its application in systems with unknown control
directions. Relaxing
these design constraints on the high-frequency gain
has consistently been a prominent research topic in the control community.

For an extended period, researchers have been dedicated to addressing this problem, 
leading to the development of numerous enhanced adaptive control methods
(\cite{lc90,rt16,ww20,gw23,jg23,wl24,lh24}).
In \cite{n83}, a Nussbaum function was first introduced into the control signal,
where the prior knowledge about the high-frequency gain is no longer required.
Since then, the adaptive control approaches on the basis of Nussbaum function have been widely reported
(\cite{yl10,zy17,lc22,hl24}). However, this method may exhibit oscillation and persistent switching issues, as shown in \cite{sk12,yh08}.
Another well-known method to addressing the issue of high-frequency gain sign 
is the multiple-model-based control technique, first developed in \cite{nb94}. 
This approach has attracted significant attention from researchers (\cite{mt16,nm21,ht22}). 
The multiple-model-based control method requires repeated estimation of plant
 parameters and may also exhibit the persistent switching issue, as mentioned in \cite{mt16}.  
 Additionally, the design constraints on the sign of the high-frequency gain can also be removed by introducing alternative design conditions, 
 such as assuming that the bounds of the high-frequency gain are known (\cite{lc90}) 
 or that the signal is interval exciting (\cite{wo23}).

Reviewing the literature reveals that researchers have made significant
long-term efforts to address the high-frequency gain sign problem,
achieving numerous remarkable results. However, some open issues
remain unresolved.
Recently, for continuous-time linear time-invariant (LTI) systems, a modified MRAC method was developed in \cite{ps22} by employing a standard Lyapunov stability
analysis.
This method relaxes the design constraints
on the high-frequency gain by injecting the estimation of the tracking
error derivative into the control signal, while mitigating the
persistent switching issue commonly involved in the well-known
Nussbaum-based and multiple-model-based methods.
Motivated by \cite{ps22}, \cite{xz24} proposed a singularity-free
output feedback MRAC method for a general class of discrete-time
LTI systems, which eliminates the need for high-frequency gain
sign and bound information commonly used in discrete-time MRAC systems. 
The development of an output feedback MRAC scheme that operates independently 
of prior knowledge concerning high-frequency gain across a general class of continuous-time  LTI systems, 
particularly those with general relative degrees, presents considerable challenges. 
These challenges stem primarily from the need to circumvent issues associated 
with persistent switching and the continual re-estimation of parameters. 
It should be emphasized that the stability analysis presented in \cite{ps22} 
relies on conventional Lyapunov techniques and is tailored specifically to MRAC systems with a relative degree of one. 
Moreover, the control methodology delineated in \cite{xz24}, although effective for discrete-time systems, 
is ill-suited for the regulation of continuous-time systems that exhibit general relative degrees. 
This incompatibility arises due to inherent differences in the stability properties between continuous and discrete-time systems. 
Consequently, there is a pronounced need for a unified, 
singularity-free output feedback MRAC framework that is adaptable to continuous-time LTI systems 
without requiring any foreknowledge of the high-frequency gain. 
This gap underscores a significant and yet unexplored area within the field of adaptive control.

In this paper, we address the challenge of adaptive control gain singularity in a general
class of continuous-time LTI systems with general relative degrees.
The fundamental strategy employed involves the decoupling of high-frequency gain
from the parameters of the derived control law. This critical decoupling
enables the transformation of the estimation error equation into a linear
regression model, distinctively diverging from the conventional bilinear
model used in traditional MRAC frameworks. Such an approach allows for
the design of the parameter update law directly, eliminating the dependency
on prior knowledge of the high-frequency gain.
The significant contributions of this research are outlined as follows:
\begin{enumerate}[(i)]
	\item We develop a novel output feedback MRAC framework applicable to 
	a comprehensive range of continuous-time LTI systems with general relative degrees.
	The proposed MRAC law is meticulously designed to ensure the boundedness of all signals within the closed-loop system, 
	as well as to achieve asymptotic output tracking of the target plant, 
	all accomplished without any reliance on prior knowledge regarding the high-frequency gain.

	\item The estimation error equation is transformed into a linear regression form, facilitated 
	by a modified MRAC framework developed in this paper. 
	This represents a significant departure from traditional MRAC systems, 
	where the estimation error equation takes on a bilinear regression form. 
	The adoption of the linear regression form enables direct estimation of all unknown parameters, 
	thereby simplifying the adaptive control process.
	
	\item A new adaptive control law with piecewise constant tuning gain is proposed to 
	address the persistent switching and repeated parameter estimation issues, 
	ensuring singularity-free operation. Unlike the primary approaches, Nussbaum and multiple-model based techniques, 
	which handle unknown high-frequency gains through persistent switching and repeated parameter estimation, 
	the proposed solution overcomes these issues despite lacking high-frequency gain information or additional design conditions.
 \end{enumerate}

The rest of this paper is structured as follows. 
In Section 2, we describe the controlled plant and clarify the technical issues to be addressed. 
In Section 3, we give the design details of the proposed adaptive control scheme. 
In Section 4, we show the simulation results. 
In Section 5, we conclude the work of this paper.

{\bf Notation:}
Let a finite-dimensional vector $X(t)  =[X_1(t), ..., X_n(t)]^T \in \mathbb{R}^n$. 
The signal space $L^2$ is defined as the set $\{X(t) : (\int_0^\infty (X_1^2(t) + \cdots + X_n^2(t)) dt)^{1/2} < \infty\}$, 
encapsulating vectors whose squared components integrated over time yield a finite result. 
Conversely, the signal space $L^\infty$ comprises vectors $X(t)$ for which $\sup_{t \geq 0} \max_{1 \leq i \leq n} |X_i(t)| < \infty$, 
indicating that each component's absolute value remains bounded over time.
For any constant $A$, the function $\operatorname{sign}(A)$ represents the sign of $A$. 
In the context of matrix algebra, for a matrix $B \in \mathbb{R}^{n \times n}$, 
the notation $B \succ 0$ signifies that $B$ is positive definite, indicating all eigenvalues are strictly positive.
The operator $s$ is employed to denote two types of operations: $sX=\mathcal{L}(X(t))$  
representing the Laplace transform of $X(t)$, or $sX = \dot X(t)$ representing the derivative of $X(t)$. 
The function $y(t) = G(s)[u](t)$ describes the output $y(t)$ of a continuous-time linear time-invariant system 
characterized by the transfer function $G(s)$, with $u(t)$ as its input. 
This notation is integral to adaptive control literature (referenced in \cite{t03, gs14, is96}) as it amalgamates operations 
in both the time domain and the complex frequency domain, 
thus obviating the necessity for complex convolutional expressions in the analysis of control systems.

\section{Problem statement}

This  section provides a description of the controlled plant 
and outlines the technical issues that need to be addressed.

\subsection{System model}

Consider a class of LTI systems, which are characterized 
by their input-output relationship in the following form:
\bea\label{s1}
P(s)[y](t) = k_p Z(s) [u](t),
\eea
where this relationship holds for all
$t\geq 0$. Here, $y(t)\in \mathbb{R}$ and $u(t) \in \mathbb{R}$ represent the output and input of the system, respectively.  
The system dynamics are governed by the polynomials  $P(s)$ and $Z(s)$,
defined as:
\bea
P(s)\ts=\ts s^{n}+p_{n-1} s^{n-1}+\cdots+p_{1} s+p_{0}\nn, \\
Z(s)\ts=\ts s^{m}+ z_{m-1} s^{m-1}+\cdots+z_{1} s+z_{0}\nn.
\eea
The coefficients $p_i $ and $z_j, 0\leq i \leq n-1, 0\leq j \leq m-1$, are unknown.
Additionally,
$k_p $, representing an unknown non-zero constant high-frequency gain, 
plays a crucial role in the system's response.
The relative degree of the system, denoted as $n^*=n-m$, 
can be arbitrary, such that $1\leq n^*\leq n$,
reflecting the differential order between the system's output and input dynamics.

The reference model in our study is defined by the following relationship:
\bea\label{r1}
y^*(t) = W^*(s)[r](t),
\eea
where $y^*(t)$ denotes the reference output, characterized by bounded derivatives up to the $n^*$ order. 
The function $W^*(s)$ represents a stable transfer function, and $r(t)$ is the reference input, 
which is also bounded along with its first derivative. 
As outlined in prior studies \cite{is96,t03},  $ W^*(s)$ is typically selected as $\frac{1}{R_m(s)}$ 
where $R_m(s)$ is a monic Hurwitz polynomial with a degree equal to   $n^*$.

\subsection{Control objective and design conditions}

This paper is dedicated to the development of a singularity-free output feedback adaptive control law 
for the system described in (\ref{s1}),  where all parameters, including $p_i$, $z_j$, and $k_p$, 
are assumed to be unknown. Our objective is to ensure the closed-loop stability of this system and 
achieve asymptotic tracking convergence, specifically that $\lim_{t\rightarrow \infty}(y(t)-y^*(t))=0$.

To advance our design and analysis, we establish the following assumptions:

{\bf(A1)}  The polynomial $Z(s)$  is Hurwitz, ensuring that it has all poles in the left half of the complex plane.

{\bf(A2)} The degree $n$ of the polynomial $P(s)$ is known.

{\bf(A3)} The system relative degree $n^*$ is known.

The closed-loop MRAC system involves zero-pole cancellation, and Assumption (A1) ensures that this cancellation is stable. 
Assumption (A2) determines the dimension of the parameters to be estimated, while Assumption (A3) pertains to the selection of $W^*(s)$.
These assumptions represent the design conditions required for the traditional MRAC framework \cite{is96,nn18,t03}.
Assumption (A2) can be relaxed if an upper bound on $n$ is known, 
and similarly, Assumption (A3) can be moderated if an upper limit on $n^*$ is established. 
These adaptations, as discussed in \cite{is96,nn18, t03}, permit a more versatile control design framework 
that accommodates uncertainties in the system's dynamic orders, 
thereby facilitating effective adaptive control performance under less rigid conditions.
Relaxing Assumptions (A2) and (A3) leads to increased complexity and additional computational demands. 
Due to the scope of this paper, we do not explore the detailed implications of these relaxations.

\subsection{Technical issues} 
\
\par {\bf Review of the traditional MRAC framework.}
To elucidate  the technical issues,
we provide a brief review of the traditional  output feedback MRAC framework.
First, we present the following lemma specifying the common matching equation in the MRAC framework.
\medskip
\begin{lemma} {\it \label{l1} (\cite{t03}) Constants $\theta_1^* \in \mathbb{R}^{n-1},\theta_2^*\in \mathbb{R}^{n-1},\theta_{3}^* \in \mathbb{R},\theta_4^* = 1/k_p$ exist such that
		\bea\label{lem11}
		\theta_1^{* T}b(s)
		{P}(s)+\left(\theta_2^{* T}b(s)+\theta_{3}^*\Omega(s)\right)k_p{Z}(s) \nn \\ =\Omega(s)\left({P}(s)-k_p\theta_4^*{Z}(s)R_m(s)\right),
		\eea
		where $b(s)=[1,s,\ldots,s^{n-2}]^T$, and $\Omega(s)$ is an any monic Hurwitz polynomial of degree $n-1$.}
\end{lemma}

\medskip
Lemma \ref{l1} formulates a key identity involving these constant vectors and polynomials, 
which plays a critical role in the stability and control analysis.
Based on Lemma \ref{l1}, the traditional output feedback MRAC law is designed as
\begin{eqnarray}\label{bscon1}
 \!\!	 u(t) \!\!  \ts=\ts   \!\! \theta_{1}^{T}(t) \phi_{1}(t) \!\!+ \!\!\theta_{2}^{T}(t) \phi_{2}(t)  \!\!+ \!\! \theta_{3}(t) y(t)  \!\!+ \!\! \theta_{4} (t) r(t),
\end{eqnarray}
where $\theta_{i}(t),i=1,2,3,4,$ are estimates of $\theta_{i}^*$, and
\bea\label{o1}
\phi_1(t)=\frac{b(s)}{\Omega(s)} [u](t),~~\phi_2(t)=\frac{b(s)}{\Omega(s)} [y](t).
\eea

Define a tracking error signal $e(t)=y(t)-y^*(t)$, and a signal $\phi(t)=[\phi_1^T(t),\phi_2^T(t),$ $y(t),r(t)]^T$. 
Let $\chi(t)$ be an estimate of $k_p$, and $\theta(t)=[ \theta_1^T(t),\theta_2^T(t),\theta_{3}(t),\theta_4(t)]^T$ be an estimate of  $\theta^{*}=[\theta_1^{* T},\theta_2^{* T},\theta_{3}^*,\theta_4^*]^T$.

Then, to estimate the parameter $\theta^*$, an estimation error is constructed as
$
\varepsilon(t) = e(t) + \chi(t) \mu(t),
$
where $\mu(t)=\theta^T(t) \varphi(t) -\frac{1}{R_m(s)} \left[\theta^T \phi\right](t)$ and $
\varphi(t)=\frac{1}{R_m(s)}\left[\phi\right](t)$.
Together with (\ref{s1})-(\ref{o1}), $\varepsilon(t)$ can be expressed as
\begin{eqnarray} \label{bs3_0}
	\varepsilon (t)= k_p\tilde \theta^T(t)\varphi(t)+\tilde \chi(t) \mu(t),
\end{eqnarray}
where $\tilde {\theta}(t) =\theta(t)-\theta^*$ and $\tilde \chi(t)=\chi(t)-k_p$.

From (\ref{bs3_0}), the parameter update law is designed as
\bea \label{bs3}
\dot{\theta}(t) \ts=\ts  -\operatorname{sign}\left(k_{p}\right) \Gamma \varepsilon(t)  \varphi(t),\nn \\
\dot{\chi }(t) \ts=\ts  -\gamma \varepsilon(t) \mu(t),
\eea
where $\Gamma\succ 0$, and $\gamma>0$ is a constant.

The derivations (\ref{lem11})-(\ref{bs3}) represent primary designs in the traditional MRAC framework. 
As discussed in \cite{is96,t03,nn18}, the MRAC law (\ref{bscon1}) with the parameter update law (\ref{bs3}) 
can guarantee closed-loop stability and $\lim_{t\rightarrow \infty}(y(t)-y^*(t))=0$.

\medskip
{\bf Clarification of the technical issues.}
In the conventional MRAC framework, the estimation error equation (\ref{bs3_0}) is characterized by a bilinear regression form. 
This configuration results in a coupling between the unknown high-frequency gain
$k_p$ and the parameter estimation error $\tilde \theta(t)$. 
Then, this interdependency complicates the direct application of standard parameter update algorithms, 
such as gradient descent and least-squares, on the parameter estimate $\theta(t)$ when $k_p$ is unknown.
Additionally, maintaining the estimate of $k_p$ as a non-zero value is crucial to prevent the MRAC law (\ref{bscon1}) from becoming singular.

To formulate a parameter estimate law that ensures the non-singularity of the MRAC law, 
most output feedback MRAC methods necessitate incorporating prior knowledge of the sign of $k_p$
into the design of the parameter update law, as   illustrated in (\ref{bs3}).
However, this requirement for prior knowledge about $k_p$ restricts the broader applicability of the MRAC methodology, 
particularly in scenarios where such information may not be readily available.

Recently, innovative strategies to remove the design dependencies on
$k_p$ were introduced in \cite{ps22,xz24}. Despite these advancements, 
as highlighted in the Introduction, these methods fall short in addressing the needs of 
a wide range of continuous-time systems with general relative degrees. To overcome the limitations of existing methods, 
particularly the issues related to persistent switching and repeated parameter estimation, 
this paper tackles several key technical challenges:
\begin{enumerate}[(i)]
	\item How to develop a new output feedback MRAC law that ensures closed-loop stability and asymptotic tracking performance,  
	while eliminating the need for prior knowledge of $k_p$ and any additional design conditions, 
	in contrast to the traditional MRAC framework?
	\item How to ensure the non-singularity of the adaptive control law throughout the parameter adaptation process, 
	in order to mitigate the complexities associated with persistent switching and frequent parameter re-estimation?
	
	\item How to  conduct a comprehensive analysis of the control performance in the closed-loop system, 
	aiming to rigorously assess the effectiveness of the proposed adaptive control methodology?
\end{enumerate}

\section{Output feedback MRAC design}

In this section, we detail the design of a novel output feedback MRAC approach tailored for the system (\ref{s1}).

\subsection{Construction of the MRAC law}

Inspired by (\ref{bscon1}), we propose the following output feedback adaptive control law:
\bea\label{con1}
u(t)= \frac{1}{1+\sigma \rho(t)} \left(\theta^T(t) \phi(t)+\sigma \theta_p^T(t) \phi(t)\right),
\eea
where $\rho(t)$ and $\theta_p(t)$ serve as   estimates of $\rho^*$ and $\theta_p^*$, respectively,
with $\rho^*=k_p$  and $\theta_p^*=k_p \theta^{*}$. Here,
$\sigma$ is employed as a piecewise constant tuning gain.
Relative to the conventional MRAC law (\ref{bscon1}), the designed MRAC formulation (\ref{con1}) 
incorporates two novel terms: $\sigma \theta_p^T(t) \phi(t)$ and $\sigma \rho(t)u(t)$.
These terms,  $\theta_p^T(t) \phi(t)$ and $\rho(t)u(t)$,
play a pivotal role in formulating the estimation error equation in a linear regression model, 
which significantly simplifies the parameter estimation process. Additionally, the term
$\sigma$ is strategically utilized to ensure that the adaptive control law (\ref{con1}) remains singularity-free. 
Further details elucidating their specific contributions and effects within the control strategy will be discussed subsequently. 
These enhancements are crucial for addressing the technical challenges previously identified.

\subsection{Specification of the new tracking error}

Operating both sides of matching equation (\ref{lem11}) on $y(t)$ results in the following expression:
\bea\label{lem1pro1}
\ts \ts \theta_1^{* T}b(s)
{P}(s)[y](t) \!\!+ \!\!\theta_2^{* T}b(s)k_p{Z}(s)[y](t) \!\!- \!\!\Omega(s){P}(s)[y](t)  \nn \\
\ts \ts \!\!= \!\! -\theta_{3}^*\Omega(s)k_p{Z}(s)[y](t) \!\!- \!\!\Omega(s) k_p\theta_4^*{Z}(s)R_m(s)[y](t).
\eea
Given the stability of $\Omega(s)$ and $Z(s)$,
and substituting (\ref{s1}) into (\ref{lem1pro1}), while disregarding terms that 
decay exponentially related to initial conditions, it can be deduced that
$
\theta_4^* R_{m}(s)[y](t)=  -\theta_{1}^{* T} \frac{b(s)}{\Omega(s)}[u](t)-\theta_{2}^{* T} \frac{b(s)}{\Omega(s)}[y](t)
-\theta_{3}^{*} y(t)+u(t).
$
Together with  (\ref{r1}) and (\ref{o1}), one obtains
\bea\label{lem1pro3}
R_{m}(s)[e](t)= \frac{-\theta^{* T} \phi(t)+u(t)}{\theta_4^*}.
\eea

It should be noted that $\theta_4^*=1/k_p$, with $\rho(t)$ and $\theta_p(t)$ 
serving as estimates for $k_p$ and $k_p \theta^*$, respectively.
The control law (\ref{con1}) can be reformulated as
\bea\label{lem1pro4}
u(t) \!\!\ts=\ts \!\!\theta^T(t) \phi(t)+\sigma \theta_p^T(t) \phi(t)-\sigma \rho(t) u(t) \nn \\
  \!\!\ts=\ts \!\! \theta^T(t)\phi(t) \!\!+ \!\!\sigma \tilde \theta_p^T(t) \phi(t)  \!\!- \!\!\sigma \tilde \rho(t) u(t) \!\!- \!\!\sigma R_m(s)[e](t).
\eea
where $\tilde \theta_p(t)=\theta_p(t)-\theta_p^{*}$ and $\tilde \rho(t)=\rho(t)-\rho^{*}$.

Let $\bar \theta(t)$ and $\lambda(t)$ represent estimates of $\bar \theta^*=[\theta^{* T},\theta_p^{* T},\rho^*]^T$ and $\lambda^*=1/k_p$, respectively. 
Substituting (\ref{lem1pro4}) into (\ref{lem1pro3}) yields that
\bea\label{lem1pro5}
\lambda^* R_m(s)[e](t) = \tilde {\bar \theta}^T(t) \omega(t) -\sigma R_{m}(s)[e](t),
\eea
where $\tilde {\bar \theta}(t)=\bar \theta(t)-\bar\theta^{* }$, $\tilde \lambda(t)=\lambda(t)-\lambda^*$ and
$\omega(t)=[\phi^T(t),\sigma \phi^T(t),-\sigma u(t)]^T$.

For parameter estimation purposes, our intention is to utilize  $R_m(s)[e](t)$
to construct an estimation error that facilitates the design of the parameter update law. 
Unfortunately, the signal
$R_m(s)[e](t)$ is not directly observable. 
To overcome this challenge, we introduce a newly defined tracking error:
\bea\label{e0}
\bar e(t) =H(s)R_m(s) [e](t),
\eea
where $H(s)$ is a designed stable filter, of the form
\bea
H(s)=\frac{1}{s^{n^*}+h_{n^*-1}s^{n^*-1}+\cdots+h_{1}s+h_0}
\eea
with $h_i,i=0,1,...,n^*-1$, being some constant parameters.
Since $P_m(s)$ and $H(s)$ have the same degree,
the signal $\bar e(t)$ is available.

Operating both sides of (\ref{lem1pro5}) by $H(s)$, one has
\bea\label{lem1pro6_1}
\left(\sigma+\lambda(t)\right) \bar e(t) = H(s)\left[\tilde {\bar \theta}^T \omega\right](t) +\tilde \lambda(t) \bar e(t).
\eea
Then, if $\sigma+\lambda(t) \neq 0$,  the new tracking error (\ref{e0}) can be expressed as
\bea\label{e2}
\bar e(t)\ts=\ts\frac{1}{\sigma+\lambda(t)} H(s)\left[\tilde {\bar \theta}^T \omega\right](t) +\tilde \lambda(t) \frac{\bar e(t)}{\sigma+\lambda(t)}
\eea
which is the desired tracking error equation and will be used for analyzing closed-loop system performance.

\subsection{Development of the parameter update law}

To formulate the parameter update law in adaptive control law (\ref{con1}), 
we introduce an estimation error signal:
\bea\label{e3}
\bar \varepsilon (t)=\bar e(t)+\frac{\eta(t)}{\sigma+\lambda(t)},
\eea
where
\bea\label{e3_1}\label{e3_2}
\eta(t)=\bar \theta^T(t) \zeta(t) -H(s) \left[\bar \theta^T \omega\right](t),~~\zeta(t)=H(s)\left[\omega\right](t).
\eea

From (\ref{e2}) and (\ref{e3}), an estimation error equation is derived as
\bea\label{e4}
\bar \varepsilon (t)=\tilde \Theta^T(t) \Phi(t),
\eea
where
\bea
\tilde \Theta(t) \ts=\ts \Theta(t)-\Theta^{*}=\left[\tilde {\bar \theta}^T(t), \tilde \lambda(t)\right]^T, \nn \\
\Phi(t)\ts=\ts\left[\frac{\zeta^T(t)}{\sigma+\lambda(t)},  \frac{\bar e(t)}{\sigma+\lambda(t)}\right]^T,
\eea
and the estimation of $\Theta^{*}=\left[\bar \theta^{* T}, \lambda^*\right]^T$ is defined as $\Theta(t)$.
Note that the derived estimation error equation (\ref{e4}) is in a linear regression form.

\begin{remark}
	The linear configuration of the estimation error equation (\ref{e4}) is derived from 
	the newly established MRAC law (\ref{con1}). 
	Analysis of (\ref{lem1pro3}) and (\ref{lem1pro4}) reveals that the term
	$\sigma \theta_p^T(t) \phi(t)-\sigma\rho(t)u(t)$
	within the new MRAC law (\ref{con1}) serves as an
	estimate of $\sigma R_{m}(s)[e](t)$. This innovative estimation approach effectively 
	allows for the decoupling of the unknown high-frequency gain 
	from the adaptive control law parameters within the estimation error equation (\ref{e4}), 
	resulting in a formulation that adheres to a linear regression model. 
	This transformation enhances the tractability  of parameter estimation within the adaptive control framework.
\hspace{\fill}$\nabla$
\end{remark}

\medskip
Based on the principles of the least-squares algorithm, we develop the following parameter update law:
\bea\label{law}
\dot \Theta(t) \ts=\ts -\frac{ \Upsilon(t)   \Phi(t) \bar \varepsilon (t)}{m^2(t)}, \nn \\
\dot \Upsilon(t)\ts=\ts - \frac{\Upsilon(t) \Phi(t) \Phi^T(t)\Upsilon(t)}{m^2(t)},
\eea
where $m(t)=\sqrt{1+\beta_1 \Phi^T(t)\Phi(t)+\beta_2 \Phi^T(t) \Upsilon(t) \Phi(t)}$
with $\beta_1$ and $\beta_2$ being positive design parameters,
and $\Upsilon(0)=\Upsilon_0 \succ 0$ is an adaptive gain matrix.

\medskip
\begin{remark}
	The estimation error equation (\ref{e4}) constructed
	in this paper differs significantly from the traditional
	MRAC framework's estimation error equation (\ref{bs3_0}),
	as it takes the form of a linear regression.
	By implementing (\ref{e4}), the parameter update law for $\Theta(t)$ can be directly designed. 
	Importantly, this novel approach negates the dependency on prior knowledge of the high-frequency gain $k_p$, 
	a staple in traditional MRAC framework. However, it is crucial to acknowledge that the elimination of $k_p$ design constraints 
	leads to an increase in the dimension of $\Theta(t)$, requiring careful consideration in implementation.\hspace{\fill}$\nabla$
\end{remark}

\medskip
The features of the parameter update law are delineated in the subsequent lemma.

\medskip
\begin{lemma}\label{lemlaw}
	{ The parameter update law (\ref{law}) ensures
		
		(i) $\Theta(t) \in L^\infty,~\frac{\bar \varepsilon (t)}{m(t)} \in L^2 \cap L^\infty ,~\dot \Theta (t) \in L^2 \cap L^\infty$; and
		
		(ii)  $\Theta(t)$ converges to a constant vector $\Theta_{\infty}$ as $t\rightarrow \infty$. }
\end{lemma}

\medskip
{\emph{Proof:}} We prove Lemma \ref{lemlaw} via the following two steps.

\medskip
\emph{Step 1:} From $\Upsilon_0 \succ 0$ and (\ref{law}),
we derive that $\frac{d}{dt}(\Upsilon^{-1}(t))\!\!=\!\!-\Upsilon^{-1}(t) \dot \Upsilon(t)\Upsilon^{-1}(t)\!\!=\frac{\Phi(t) \Phi^T(t)}{m^2(t)}$.
It can be seen that $\Upsilon^{-1}(t)$ is nondecreasing.
Operating both sides of the above equation on integral yields
\bea\label{lempro0}
\Upsilon^{-1}(t)=\Upsilon^{-1}_0+\int_{0}^{t} \frac{\Phi(\tau)\Phi^T(\tau)}{m^2(\tau)} d \tau.
\eea
Note that $\Upsilon_0 \succ 0$ and $\int_{0}^{t} \frac{\Phi(\tau)\Phi^T(\tau)}{m^2(\tau)} d \tau \geq 0$.
Then, it follows that  $\Upsilon^{-1}(t)$ is also positive definite for any $t \geq 0$.
Together with (\ref{law}), one has that $\dot \Upsilon(t)$ is bounded.

Introduce a Lyapunov function candidate $V_\Upsilon=\tilde \Theta^T(t) \Upsilon^{-1}(t) \tilde \Theta(t)$, we have
\bea
\dot V_\Upsilon=-\frac{\tilde \Theta^T(t) \Phi(t)\Phi^T(t)\tilde \Theta(t)}{m^2(t)}=-\frac{\bar \varepsilon^2 (t)}{m^2(t)} \leq 0,
\eea
which implies that $V_\Upsilon$ is convergent and
bounded. Therefore, $\frac{\bar \varepsilon(t)}{m(t)} \in L^2$.
Substituting (\ref{lempro0}) into the Lyapunov function candidate $V_\Upsilon$, one has
\bea\label{lempro2}
V_\Upsilon=\tilde \Theta^T(t) \Upsilon^{-1}_0\tilde \Theta(t)+\tilde \Theta^T(t) \int_0^t \frac{\Phi(\tau)\Phi^T(\tau)}{m^2(\tau)} d \tau \tilde  \Theta(t).
\eea

It follows from (\ref{lempro2}), the boundedness of $V_\Upsilon$ and the fact $\Upsilon^{-1}_0 \succ 0$ yield that  $\tilde \Theta^T(t) \Upsilon^{-1}_0\tilde \Theta(t)\in L^\infty$, as well as $\tilde \Theta(t)\in L^\infty$ and $\Theta(t)\in L^\infty$.
Then, from (\ref{e4}), we deduce the boundedness of $\frac{\bar \varepsilon(t)}{m(t)}$.
Besides, based on (\ref{law}), one has
\bea\label{lempro3}
\|\dot \Theta(t)\| \leq \frac{\|\Upsilon^{\frac{1}{2}}(t)\| \|\Upsilon^{\frac{1}{2}}(t) \Phi(t)\|}{\sqrt{1+\beta_1 \Phi^T(t)\Phi(t)+\beta_2 \|\Upsilon^{\frac{1}{2}}(t) \Phi(t)\|^2}} \frac{|\bar \varepsilon(t)|}{m(t)},
\eea
which implies that $\dot \Theta(t) \in L^\infty$.
With $\frac{\bar \varepsilon(t)}{m(t)} \in L^2$ and (\ref{lempro3}),
we have $\dot \Theta(t) \in L^2$.

\medskip
\emph{Step 2:} From $\Upsilon^{-1}(t)\succ 0$ and (\ref{law}), one has $\Upsilon(t)\!=\!\Upsilon_0-\int_{0}^{t} \frac{\Upsilon(\tau)\Phi(\tau)\Phi^T(\tau)\Upsilon(\tau)}{m^2(\tau)} d \tau \succ 0$.
Thus, for any constant vector $z$ with an appropriate dimension, the following
formula holds
\bea
z^T \Upsilon(t) z =z^T \Upsilon_0 z -f(t) \geq 0,
\eea
where $f(t)=\int_0^t \frac{z^T \Upsilon(\tau)\Phi(\tau)\Phi^T(\tau)\Upsilon(\tau)z}{m^2(\tau)} d \tau$.
Further, considering that $f(t)$ is nondecreasing and $f(t)\leq z^T \Upsilon_0 z$,
it follows from there are constants $\varpi_1$ and $\varpi_2 \geq 0$ such that $\lim_{t \to \infty}f(t)=\varpi_1$ and $\lim_{t \to \infty}z^T \Upsilon(t) z=\varpi_2$.
Then, the arbitrariness of the constant vector $z$ implies
$ \Upsilon(t)$ converges to some constant matrix
$\Upsilon_{\infty}$.

Based on (\ref{law}), we obtain $\dot {\tilde \Theta}(t)=\dot \Upsilon(t) \Upsilon^{-1}(t) \tilde \Theta(t)$ 
whose solution is $\tilde \Theta(t)=\Upsilon(t) \Upsilon^{-1}_0 \tilde \Theta(0)$.
Together with $\lim_{t \to \infty}\Upsilon(t)=\Upsilon_{\infty}$, we conclude that
$
\lim_{t \to \infty} \Theta(t)=\Theta^*+\lim_{t \to \infty}\Upsilon(t)\Upsilon^{-1}_0 \tilde \Theta(0)=\Theta_{\infty}.
$
\hspace{\fill}$\square$

\subsection{Design of the tuning gain}

To guarantee the non-singularity of the adaptive control law $u(t)$ specified and to 
ensure the valid formulation of the tracking error equation  $\bar e(t)$ given in (\ref{e2}), 
it is crucial that the tuning gain $\sigma$ is selected such that 
it meets the following conditions for all feasible values of  $\rho(t)$ and $\lambda(t)$:
\bea\label{g0_1}
1+\sigma \rho(t) \ts\neq\ts 0, \\
\label{g0_2}
\sigma+\lambda(t) \ts\neq\ts 0.
\eea

To this end, the tuning gain $\sigma$ is designed as
\bea\label{g1}
\sigma=\begin{cases} 1 & \text {if } \varsigma(t)\geq 1 \text{ or } \rho(t)=\lambda(t)= 0,  \\
	-1 & \text {if } \varsigma(t)\leq -1,
	\\ 0 & \text {otherwise},\end{cases}
\eea
where $\varsigma(t)=\operatorname{sign}\left(\rho(t)\right)+\operatorname{sign}\left(\lambda(t) \right)$.
It is evident that the tuning gain obtained from (\ref{g1}) ensures
$1+\sigma \rho(t) \geq 1$ and $\sigma+\lambda(t) \neq 0$,
and thus (\ref{g0_1}) and (\ref{g0_2}) hold.
As established by Lemma \ref{lemlaw},  $\rho (t)$  and $\lambda(t)$ asymptotically converge to their respective constant values.
This convergence implies that the tuning gain $\sigma$
necessitates only a limited number of adjustments throughout the time interval
$[0,\infty)$.
Consequently, the adaptive control law (\ref{con1}) circumvents the persistent switching problem 
that is often encountered in control methodologies reliant on Nussbaum functions and multiple-model approaches. 
Furthermore, condition (\ref{g1}) ensures that $1/(1+\sigma \rho(t)) \leq 1$, 
signifying that the proposed adaptive control strategy effectively avoids issues associated with excessive gain magnitudes.

\medskip
\begin{remark}
	{\rm The selection of values for
		$\sigma$ offers flexibility and is not constrained to a single value.
		Specifically, the value of $\sigma$ in (\ref{g1}) can also be chosen as any positive constant given 
		that $\varsigma(t)\geq 1$ or  $\rho(t)=\lambda(t)=0$. Conversely,
		$\sigma$ can be set to any negative constant when $\varsigma(t)\leq -1$.
		These adjustments are crucial as they ensure compliance with conditions (\ref{g0_1}) and (\ref{g0_2}).
	}\hspace\fill$\nabla$
\end{remark}

\subsection{Stability analysis}

Based on the derivations presented, the key findings of this paper are outlined as follows.

\medskip
\begin{theorem}\label{thm}{\it
		Assuming that Assumptions (A1)-(A3) are satisfied for the system described in (\ref{s1}), 
		the implementation of the adaptive control law (\ref{con1}), 
		coupled with the parameter update law (\ref{law}) and the tuning gain defined in (\ref{g1}), 
		guarantees the stability of the closed-loop system and ensures that $\lim_{t\rightarrow \infty}\left(y(t)-y^*(t)\right)=0$.
	}	
\end{theorem}

\medskip
{\emph{Proof:}}
The proof consists of three steps. This proof utilizes some lemmas and notation, including Lemma \ref{lemA}, Lemma \ref{lemB}, 
$c,\kappa,\varrho ,L^{\infty e}$ and $\|\cdot\|_t$, which are presented in  Appendix A. 
To conserve space, the exponentially decaying terms associated with  initial conditions are omitted.

\medskip
\emph{Step 1: Demonstrate that    $\|\omega(t)\|_t\leq c \|\tilde {\bar \theta}^T(t)\omega(t)\|_t+c$.}

Given that $Z(s)$ is a Hurwitz polynomial,  the control input
$u(t)$ is defined as
\bea
u(t)=\frac{1}{k_p} \frac{F(s)D(s)}{Z(s)}[y](t)=\frac{1}{k_p} \frac{F(s)}{Z(s)}[D(s)[y]](t),
\eea
where $F(s)D(s)=P(s)$ with $F(s)$ and $D(s)$ as two polynomials of degree $m$ and $n^*$, respectively.
The transfer function $\frac{F(s)}{Z(s)}$ is stable and proper, leading to the following bound on $u(t)$:
\bea\label{thmpro3}
\|u(t)\|_t
\leq  c\|y(t)\|_t +  c\|\dot y(t)\|_t + \cdots + c\|y^{(n^*)}(t)\|_t +c.
\eea

Referencing (\ref{r1}), (\ref{lem1pro5}), and the stability of $R_m(s)$, we establish
\bea\label{thmpro5}
\|y^{(i)}(t)\|_t \leq \|e^{(i)}(t)\|_t+c \leq c \|\tilde  {\bar \theta}^T(t) \omega(t)\|_t+c,
\eea
for $i=0,1,2,\ldots,n^*$.
It follows from (\ref{o1}) and $\omega(t)=[\phi^T(t),\sigma \phi^T(t),-\sigma u(t)]^T$ that
\bea\label{thmpro5_1}
\|\omega(t)\|_t \leq c \|u(t)\|_t+c \|y(t)\|_t+c.
\eea

Then, combining (\ref{thmpro3})-(\ref{thmpro5_1}) yields that
\bea\label{thmpro6}
\|u(t)\|_t \ts\leq\ts c \| \tilde  {\bar \theta}^T(t) \omega(t)\|_t+c, \\
\label{thmpro7}
\|\omega(t)\|_t \ts\leq\ts c \| \tilde  {\bar \theta}^T(t) \omega(t)\|_t+c.
\eea

\medskip
\emph{Step 2: Show that $\|\frac{d }{dt}(\tilde {\bar\theta}^T(t)\omega(t))\|_t\leq \|\tilde {\bar\theta}^T(t)\omega(t)\|_t+c$.}

From Lemma \ref{lemlaw} and the derivative expression:
\bea
\frac{d }{dt}(\tilde {\bar\theta}^T(t)\omega(t))=\dot {\bar\theta}^T(t) \omega(t)+\tilde {\bar\theta}^T(t) \dot\omega(t),\eea
it follows that
\bea
\|\frac{d }{dt}(\tilde {\bar\theta}^T(t)\omega(t))\|_t \leq c \|\omega(t)\|_t+c \|\dot \omega(t)\|_t+c.\eea
Together with (\ref{thmpro7}), we have
\bea\label{thmpro13}
\|\frac{d }{dt}(\tilde {\bar\theta}^T(t)\omega(t))\|_t \leq c \|\tilde  {\bar\theta}^T(t) \omega(t)\|_t+c \|\dot \omega(t)\|_t+c.
\eea

Given the stability of $\frac{b(s)}{\Omega(s)}$ and prior equations  (\ref{r1}), (\ref{o1}) and (\ref{con1}), we obtain that
$
\|\dot u(t)\|_t  \leq  c \| \phi(t)\|_t + c \| \dot \phi(t)\|_t + c$,
$\|\dot\phi(t)\|_t  \leq  c \|u(t)\|_t + c \| y(t)\|_t + c \|\dot y(t)\|_t  + c \|\dot r(t)\|_t + c$ and
$\|\dot\omega(t)\|_t  \leq c \| \dot \phi(t)\|_t + c \| \dot u(t)\|_t + c.
$
Together with (\ref{thmpro5}) and (\ref{thmpro6}),
{$\|\dot\omega(t)\|_t \leq c \| \tilde  {\bar\theta}^T(t) \omega(t)\|_t+c$ can be obtained.}
Then, from (\ref{thmpro13}), one has
\bea\label{thmpro14}
\|\frac{d }{dt}(\tilde {\bar\theta}^T(t)\omega(t))\|_t \leq c \|\tilde  {\bar\theta}^T(t) \omega(t)\|_t+c.
\eea

\medskip
\emph{Step 3: Show the boundedness of all signals within the closed-loop system and the convergence of $ y(t)-y^*(t) $ to zero as $t\rightarrow \infty$.}

Since $H(s)$ is a stable filter {with} $n^*$ and $R_m(s)$ is a Hurwitz polynomial {with $n^*$,
	one can deduce} from (\ref{e0}) that $\|\bar e(t)\|_t \leq c \| e(t) \|_t+c$.
Based on (\ref{e4}) and (\ref{law}), we have that $\frac{d}{dt}(\frac{\bar \varepsilon(t)}{m(t)}) \in L^\infty$.
Then, it follows from $\frac{\bar \varepsilon(t)}{m(t)} \in L^2 \cap L^\infty $ that
\beq\label{thmpro15}
\|\bar\varepsilon(t)\| \leq \frac{\|\bar\varepsilon(t)\|}{m(t)}(1+c\|\zeta(t)\|+c\|\bar e(t)\|)
\leq \varrho  (1+\|\zeta(t)\|+\|\bar e(t)\|).
\eeq
From (\ref{e3_2}), (\ref{thmpro5}), (\ref{thmpro7}), (\ref{thmpro15}) and $\|\bar e(t)\|_t \leq c \| e(t) \|_t+c$, we derive
\bea\label{thmpro16}
\|\bar \varepsilon(t)\|_t \leq  \varrho   + \varrho  \|\omega(t)\|_t  +  \varrho  \|\bar e(t)\|_t \leq  \varrho   + \varrho  \|\tilde   {\bar\theta}^T(t) \omega(t)\|_t.
\eea

By invoking Lemma \ref{lemA} in Appendix, $\eta(t)$ in (\ref{e3_1}) can be expressed as
\beq
\eta(t) \!\! = \!\! {\bar \theta}^T(t) H(s)[\omega ](t) -H(s) [{\bar \theta}^T \omega](t)
 \!\!= \!\!  H(s) [ H(s) [\omega^T] \dot {\bar \theta} ](t). \nn
\eeq
Then, according to  $\dot \Theta(t) \in L^2 \cap L^\infty$, one has that
$
\| H(s) [\omega^T](t) \dot {\bar \theta}(t)\|_t \leq \kappa \|\omega(t)\|_t + \kappa.
$
Furthermore, from Lemma \ref{lemB} in Appendix and (\ref{thmpro6}), we obtain
\bea\label{thmpro19}
\|\eta(t)\|_t  \ts\leq\ts \varrho  \|\omega(t) \|_t+\varrho  \leq\varrho  \|\tilde  {\bar \theta}^T(t) \omega(t)\|_t+\varrho .
\eea
Thus, it follows from (\ref{e3}), (\ref{thmpro16}), (\ref{thmpro19}) and $\Theta(t) \in L^\infty$ that
\bea\label{thmpro20}
\|\bar e(t)\|_t \leq \|\bar \varepsilon(t) \|_t+c\|\eta(t)\|_t \leq \varrho + \varrho  \|\tilde  {\bar \theta}^T(t) \omega(t)\|_t.
\eea

From Lemma \ref{lemB} in Appendix, (\ref{lem1pro6_1}) and (\ref{thmpro14}), we can derive
\bea\label{thmpro21}
\|\tilde  {\bar \theta}^T(t) \omega(t)\|_t \leq c\|\bar e(t)\|_t +c.
\eea
Substituting (\ref{thmpro21}) into (\ref{thmpro20}), it yields that
$
\|\bar e(t)\|_t \leq \varrho + \varrho  \|\bar e(t)\|_t$, {which also implies the existence of $\bar e(t)$ boundary as well as} $\lim_{t\rightarrow \infty}e(t)=0$.
Together with (\ref{thmpro21}) and $\tilde \Theta(t) \in L^\infty$, 
one has that $\omega(t)$ is bounded. Consequently, the boundedness of $\phi(t)$ and $u(t)$ can also be obtained.
\hspace{\fill}$\square$

\medskip
Thus far, we have introduced an enhanced output feedback MRAC scheme tailored 
for continuous-time LTI systems with general relative degrees in our analysis.  
This scheme eliminates the requirement for prior knowledge of the high-frequency gain and additional design constraints, 
simplifying the implementation and broadening the applicability of the control strategy.
\medskip
\begin{remark}
	{\rm
		Different from the literature \cite{ps22} and \cite{xz24},
		this paper proposed a unified MRAC solution for continuous-time LTI systems with a general relative degree of $n^*$ ($1\leq n^*\leq n$).
		The control framework comprises the adaptive control law (\ref{con1}), the new tracking error (\ref{e0}), 
		the parameter update law (\ref{law}), and the tuning gain (\ref{g1}). 
		These components are used to analyze the control performance based on signal input-output properties.
		Moreover, the proposed MRAC solution avoids the issues of singularity, persistent switching, repeated parameter estimation and high-gain.
	}\hspace\fill$\nabla$
\end{remark}

\section{Simulation Example}
In this section, the proposed adaptive control scheme is verified through an aircraft control system.

\medskip
{\bf Aircraft model.} Consider a linearized longitudinal dynamics model of the Boeing 737 as follows \cite{fp02,t04}
\bea\label{sim1}
P(s)[\vartheta](t)=-0.023 Z(s)[\delta](t),~t\geq 0,
\eea
where $\vartheta$ is the pitch angle, $\delta$ is the deviation of the elevator,
and the high-frequency gain is $k_p=-0.023$.
Here,
\bea
P(s)\ts=\ts s^4+1.379s^3+2.174s^2+0.989s+0.065, \nn \\
Z(s)\ts=\ts s^2+0.767s+0.050, \nn
\eea
where $Z(s)$ is stable of degree two, $P(s)$ is of order $n=4$, and {$n^*=2$ is denotes 
	as the system relative degree of this simulation example}.

In this simulation, the output signal is $y=\vartheta$ and the input signal is $u=\delta$.
It's assumed that none of the components for  $Z(s),P(s)$ and $k_p$ are known. Moreover, $n$ and $n^*$ are known. 
The reference output is {selected} as
$
y^*(t)=\frac{1}{R_m(s)}[r](t),
$
where $R_m(s)=s^2+21s+108$ and $r(t)=\sin t-0.5 \sin 0.5t$.

{\bf Control law parameter calculation.}
Based on (\ref{lem11}) and (\ref{sim1}), we calculate
\bea
\theta_1^*\ts=\ts[9.856,-2.987,-20.388]^T, \nn \\
\theta_2^*\ts=\ts[-71588.696, -105840.673,-36294.042]^T, \nn \\
\theta_{3}^*\ts=\ts 11059.088, ~~\theta_4^*=1/k_p=-43.478, \nn \\
\theta_p^*\ts=\ts k_p\theta^*=[-0.226,0.069,0.468, \nn \\
\ts\ts 1646.540,2434.335,834.763,-254.359,1]^T, \nn \\
\rho^*\ts=\ts k_p=-0.023,~~\lambda^*=1/k_p=-43.478.\nn
\eea

Let $\theta_{3}(t),\theta_{4}(t),\rho(t),\lambda(t)$ and $\Theta(t)$ be estimates of $\theta_{3}^*,\theta_{4}^*,\rho^{*},\lambda^*$ and {$\Theta^*$, respectively, 
	in which the specific expression for $\Theta^*$ is $\Theta^*=[\theta_1^{* T},\theta_2^{* T},\theta_{3}^*,\theta_{4}^*,\theta_p^{* T},\rho^{*},\lambda^*]^T$.} 
	Then, we apply the MRAC law (\ref{con1}) with the parameter update law (\ref{law}) and the tuning gain (\ref{g1}) to the simulation model (\ref{sim1}),
where $\Omega(s)=s^3+8s^2+18.25s+11.25,h_1=21$ and $h_0=108$.
Here, we validate the proposed adaptive control scheme through two cases.
(i) The initial estimation for $\sign(k_p)$ is correct, and the initial value for $\Theta(t)$ is set as $\Theta(0)=[1.2\theta_1^{* T},1.2\theta_2^{* T},1.2\theta_{3}^*,1.2 \theta_4^*,0.9\theta_p^{* T},1.2\rho^{*},0.8\lambda^*]^T$.
(ii) The initial estimation for $\sign(k_p)$ is wrong, and the initial value for $\Theta(t)$ is set as $\Theta(0)=[0.8\theta_1^{* T},0.8\theta_2^{* T},0.8\theta_{3}^*,-0.3\theta_4^*,-0.5\theta_p^{* T},-0.5\rho^{*},-0.5\lambda^*]^T$.

\begin{figure}[!ht]
		\vspace{-0.3cm}
	\includegraphics[height=3.7 cm, width=8cm]{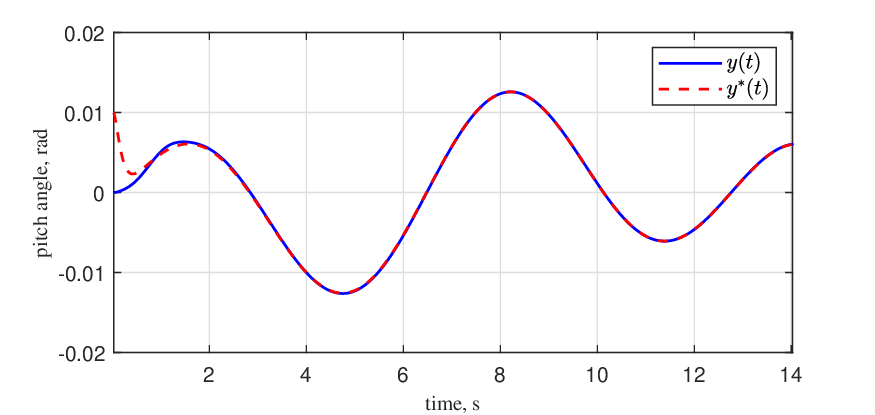}
	\centering
		\vspace{-0.3cm}
	\caption{{Responses of} the output $y(t)$ and $y^*(t)$ (Case (i)).}  \label{f1}
\end{figure}

\begin{figure}[!ht]
		\vspace{-0.5cm}
	\includegraphics[height=5.5 cm, width=8cm]{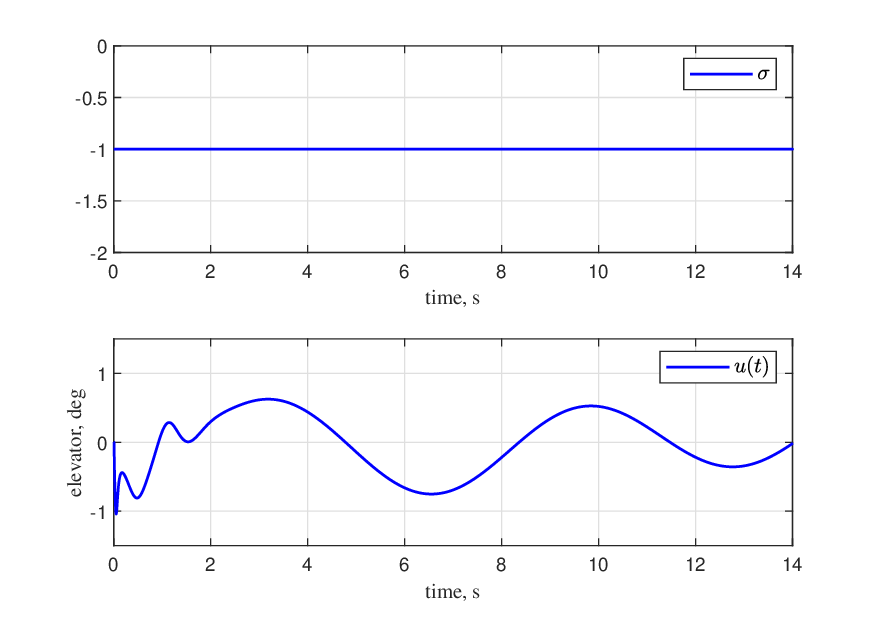}
	\centering
		\vspace{-0.3cm}
	\caption{{Evolutions of} the tuning gain $\sigma$ and input $u(t)$ (Case (i)).}  \label{f2}
\end{figure}

\begin{figure}[!ht]
		\vspace{-0.5cm}
	\includegraphics[height=7.5 cm, width=8cm]{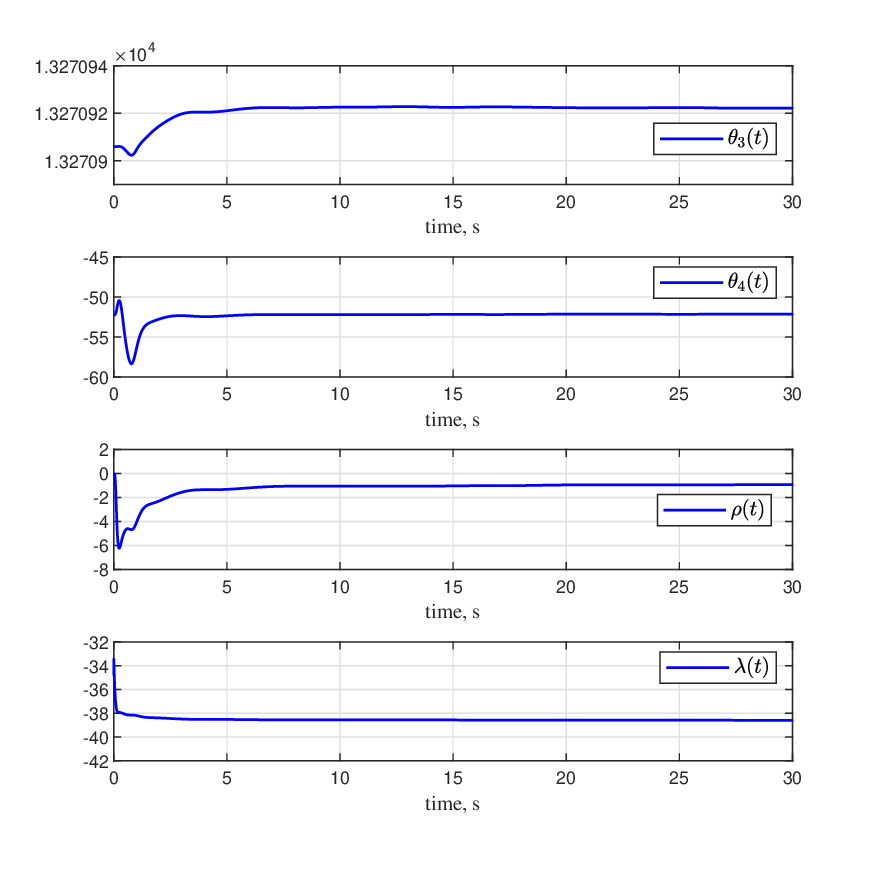}
	\centering
		\vspace{-0.3cm}
	\caption{{Evolutions of} part of parameters in $\Theta(t)$ (Case (i)).}  \label{f3}
\end{figure}

\medskip
{\bf Simulation responses.} For Case (i), the trajectories of the system output and reference output are shown in Fig. \ref{f1}.
The trajectories of the tuning gain $\sigma$ and input $u(t)$ are depicted in Fig. \ref{f2}.
The adaptation of a subset of parameters in $\Theta(t)$ is plotted in Fig. \ref{f3}.
It is evident
that the sign of $\rho(t)$ and $\lambda(t)$ remains correct and unchanged
during the process of parameter adaptation,
leading to $\sigma$ also being unchanged.

\begin{figure}[!ht]
	\vspace{-0.3cm}
	\includegraphics[height=3.7 cm, width=8cm]{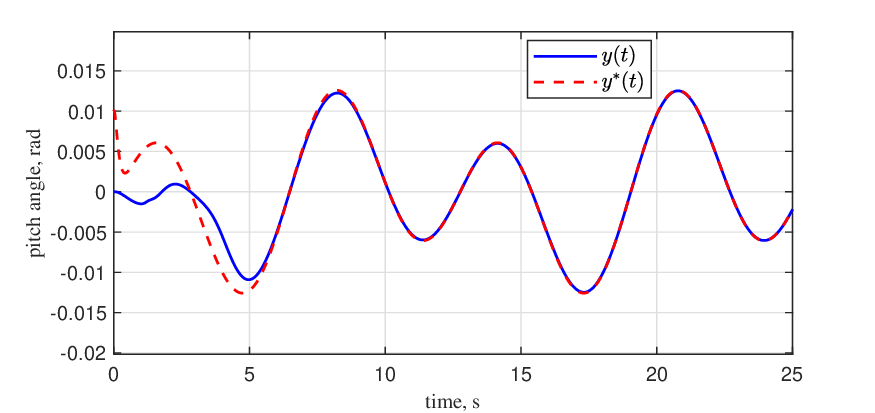}
	\centering
	\vspace{-0.3cm}
	\caption{{Responses of} the output $y(t)$ and $y^*(t)$ (Case (ii)).}  \label{f4}
\end{figure}

\begin{figure}[!ht]
	\vspace{-0.5cm}
	\includegraphics[height=5.5 cm, width=8cm]{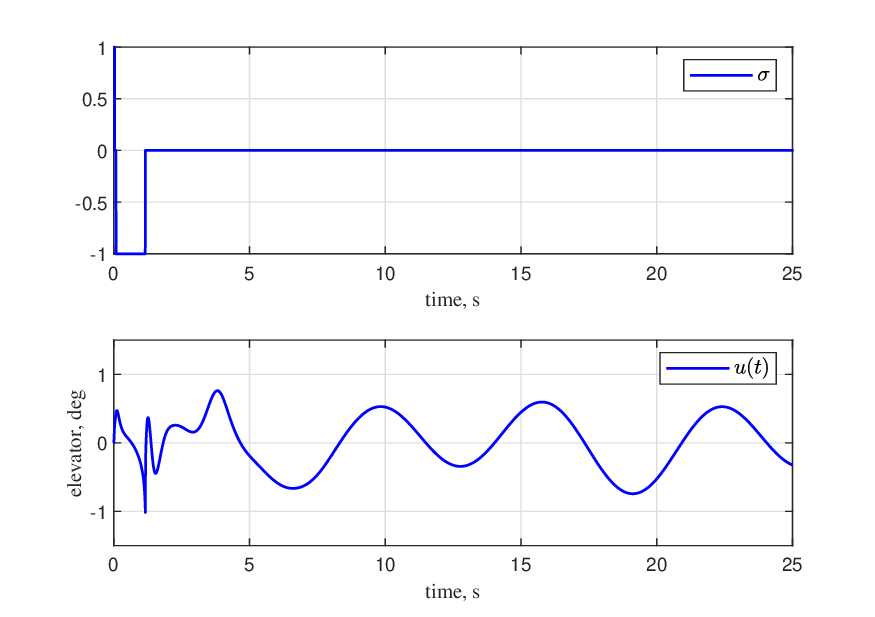}
	\centering
	\vspace{-0.4cm}
	\caption{{Evolutions of} the tuning gain $\sigma$ and input $u(t)$ (Case (ii)).}  \label{f5}
\end{figure}

\begin{figure}[!ht]
	\vspace{-0.3cm}
	\includegraphics[height=7.5 cm, width=8cm]{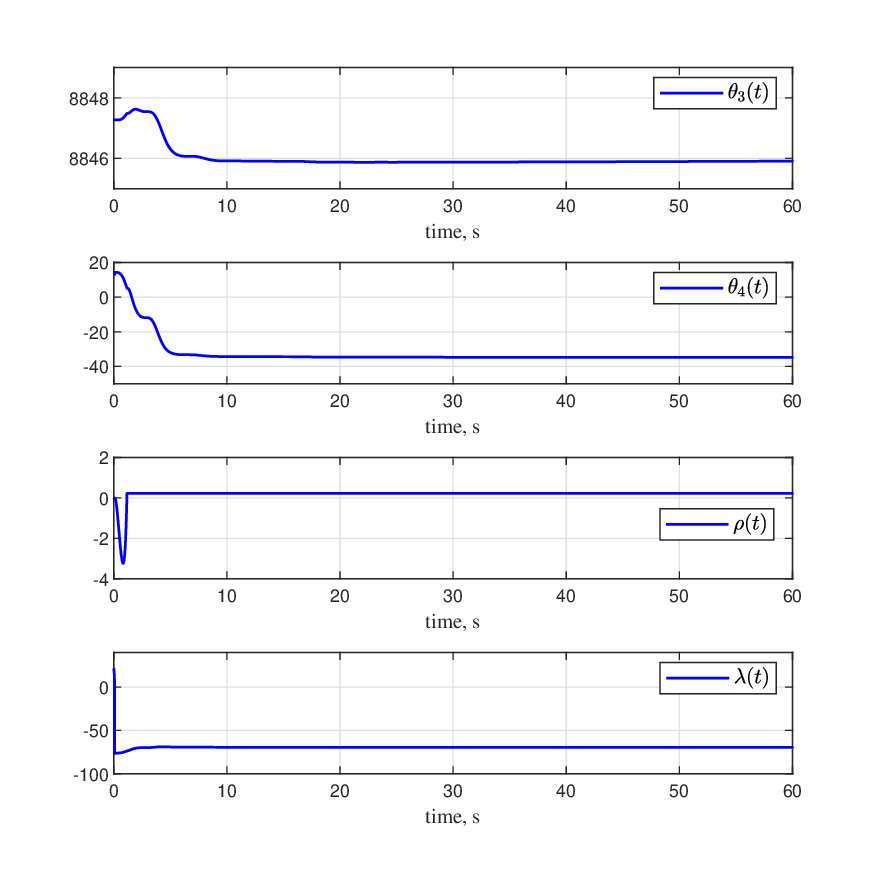}
	\centering
	\vspace{-0.6cm}
	\caption{Evolutions of parameters in $\Theta(t)$ (Case (ii)).} \label{f6}
\end{figure}

For Case (ii), the response of the system output is shown in Fig. \ref{f4}.
The {evolutions of} the tuning gain $\sigma$ and input $u(t)$ is depicted in Fig. \ref{f5}.
The {evolutions of} a subset of parameters in $\Theta(t)$ is shown in Fig. \ref{f6}.
As can be seen from Figs. \ref{f1} and \ref{f4}, compared with that of Case(i),
the transient response of the output tracking of Case (ii) has a longer rise time and
larger overshoot.
This phenomenon is primarily due to the incorrect initial estimation of the sign of $k_p$.
Recall that $\theta_{4}^*=1/k_p$, $\lambda^*=1/k_p$, and $\rho^*= {k_p}$, 
with their signs being consistent with that of $k_p$. However, as depicted in Fig. \ref{f6}, 
the signs of $\theta_{4}(t)$, $\lambda(t)$, and $\rho(t)$ may not be consistent with that of $k_p$.
Nevertheless, the proposed adaptive control scheme can still meet the control
objective.  The incorrect estimation of the sign of $k_p$ can be compensated by automatically adjusting the tuning gain $\sigma$.
For instance, even if the sign of $\rho(t)$ is positive, which is opposite to that of $k_p$,
any potential negative impact on the closed-loop system is avoided by automatically
setting $\sigma$ to zero.

The above results demonstrate that the adaptive control scheme proposed in this paper
ensures system stability and asymptotic output tracking
without requiring any prior knowledge of the high-frequency gain.
Meanwhile, the adaptive control law exhibits no singularity or high-gain issues.
Moreover, the parameter estimates converge to some constants,
and thus the persistent switching issue is avoided.

\section{Conclusion}
This paper presents a new singularity-free MRAC scheme with piecewise constant tuning gain, 
addressing the unknown high-frequency gain problem in spite of lacking any prior knowledge about high-frequency gain or additional design conditions. The proposed scheme avoids the drawbacks of Nussbaum and multiple-model approaches, which necessitate persistent switching or repeated parameter estimation to handle unknown high-frequency gains, while guaranteeing stability 
and achieving asymptotic output tracking of the closed-loop system.
Specifically, our scheme transforms the estimation error equation 
into a linear regression form, enabled by a modified MRAC law we developed.
This is a key departure from traditional MRAC systems,
where the equation for estimating the error takes the form of a bilinear regression.
The utilizing of linear regression form simplifies the adaptive control process, 
which allows for the direct estimation of all unknown parameters. Meanwhile, our scheme designs a piecewise constant tuning gain driven by the least-squares-based parameter update law. This tuning gain ensures that a modified MRAC law we developed achieves the desired control performance with a limited number of adjustments. Finally, our scheme is applied to an aircraft control system, and simulation results demonstrate its effectiveness.
Future studies will further extend the proposed adaptive control scheme to multi-input and multi-output systems 
and relax the design constraints on the unknown high-frequency gain matrix.

\appendix
\section{Two useful lemmas}

The following section introduces  two lemmas utilized in the proof of Theorem \ref{thm}.

\medskip
\begin{lemma} (\cite{sb11}) \label{lemA} {\it For any vectors $\varpi (t) \in \mathbb{R}^n$ and $\xi(t) \in \mathbb{R}^n$, if $G_m(s)=c_m(sI-A)^{-1} b_m$  is the minimal realization of a proper transfer function,
		the following equation holds
		\bea
			\varpi^T(t) G_m(s)[\xi](t)-G_m(s)[\varpi ^T \xi](t)= \nn \\
			c_m(sI-A)^{-1}[(sI-A)^{-1}b_m[\xi ^T](t) \dot{\varpi }](t),
		\eea
		where $\dot \varpi (t)$ is the first-time derivative of $\varpi (t)$.}
\end{lemma}

\medskip
\begin{lemma} (\cite{sb11}) \label{lemB} {\it Define some notation: $c$ denotes a signal bound;
		$\kappa$ and $\varrho$ denote generic $L^2 \cap L^\infty$ functions with $\varrho $ asymptotically approaching zero   as $t \to \infty$;
		$\|x\|_t$ denotes the norm $\sup_{q \leq t} \max_{1 \leq i \leq n} |x_i(q)|$ with $x(t)=[x_1(t),...,x_n(t)]^T \in \mathbb{R}^n$ being  a finite-dimensional vector;
		and
		$L^{\infty e}$ denotes the set $\{g(t)~|~ \forall d < \infty,g_d(t) \in L^\infty\}$, for any function $g(t)$ with $g_d(t)=g(t), t \leq d$, and otherwise, $g_d(t)=0$.
		Then, the following results hold
		\begin{itemize}
			\item[(i)] For a system characterized by $y(t)=G(s)[u](t)$ where  $G(s)$ is a strictly proper and stable transfer function, the inequality
			$ \|y\|_t \leq \varrho \|v\|_t+\varrho $ is satisfied if  $\|u\|_t \leq \kappa\|v\|_t+\kappa$.
			\item[(ii)] For a system characterized by  $y(t)=G(s)[u](t)$ where $G(s)$  is a proper and minimum phase transfer function,
			the inequality $\|u\|_{t} \leq c\|y\|_{t}+c$ is satisfied
			if  $u$ and $\dot{u}$ belong to $L^{\infty e}$, and  $\|\dot{u}\|_{t} \leq c\|u\|_{t}+c$.
	\end{itemize}}
\end{lemma}

\clearpage
\end{document}